RESEARCH ARTICLE

# Let Cyclic Electrochemical Data Speak for Your Energy Storage Material and Processing

Vinod Sarky[1,2] | P Laxman Mani Kanta[3] | Shivangi Keshri[1,2] | Mannanvali Shaik[1,2] | B R K Nanda[2,4] | Satyesh K. Yadav[1,2]

[1]Department of Metallurgical and Materials Engineering, Indian Institute of Technology Madras, Chennai, 600036, Tamil Nadu, India | [2]Center for Atomistic Modeling and Materials Design, Indian Institute of Technology Madras, Chennai, 600036, Tamil Nadu, India | [3]Department of Metallurgical and Materials Engineering, National Institute of Technology Tiruchirappalli, Tiruchirappalli, 620015, Tamil Nadu, India | [4]Department of Physics, Indian Institute of Technology Madras, Chennai, 600036, Tamil Nadu, India

**Correspondence:** Satyesh K. Yadav (satyesh@iitm.ac.in)

## ABSTRACT

Efforts to improve performance, like energy and power density of electrochemical energy storage devices (batteries, capacitors, and super-capacitors), are being made using a range of experimental and computational tools. The most common strategy involves exploring new material chemistries, developing advanced synthesis routes, and advanced ways to integrate these materials into device architectures. Although it is desirable to evaluate the performance of such efforts at a fixed cell format, not all studies conduct testing across the same cell formats. Specific performance normalized per mass of active material has been adopted as a universal indicator of performance, but it is not a reliable way to compare across cell formats. We propose a new mass normalization, i.e., the mass of electrochemical participants, to calculate specific performance. This new specific performance makes it possible to efficiently compare the efficacy of materials, synthesis, and cell assembly across different cell formats. We developed a framework to comprehensively report materials, synthesis, and cell assembly, and performance data, which facilitates succinctly reporting all performance indicators, including charge discharge curves over all cyclic stability assessments. The new specific performance and mass of the electrochemical participants per unit area of the electrode allows us to create a calculator to estimate performance at any commercial device-level cell format from measured data at the laboratory-level cell. The new specific performance also helps in optimizing the coating thickness and the loading of the active material. The framework will facilitate reliable screening of materials and methods that could be suitable for commercial adoption. The framework can be adapted by journals to facilitate the reporting of data in a comprehensive way. Leveraging this framework, we have also compiled a database (hosted at https://power.tattvasar.com/) by extracting pertinent data from existing literature.

## 1 | Introduction

There are several efforts reported every day on improving the performance of electrochemical energy storage devices by exploring a wide array of material chemistries for active materials, electrolyte, separator, advanced synthesis techniques, and assembly strategies to build a cell. These devices, like intercalation-type devices (primary and secondary electrical batteries, solid-state batteries etc.), capacitors, and supercapacitors, share certain fundamental features namely, i) the cathode, the anode, the separator (or dielectric in the case of capacitors), the electrolyte, and the metal current collectors, ii) the principle of using potential difference between two electrodes to deliver the current. Therefore, these commonalities must be leveraged to compare the performance of all these devices within a common framework. However, different standards of reporting followed for capacitors, supercapacitors, and intercalation-type devices hinder the process of comparison. It is also not possible to rank the electrochemical performance reported across various reports by comparing the specific performance normalized by the total mass of the cell, which is mostly preferred. It is because the research community mostly reports the specific performance normalized to mass of active material, the only comparison that can be reliably made across various reports is using it. This specific performance, normalized only by the mass of active material, ignores the role of other components in a cell - such as passive material, electrolyte, separator, and current collectors, which also contribute to the performance of a cell. As a result, reliable conclusions about a cell's performance could not be made. The physical dimensions such as the area of the electrode and the masses of various components of the cell, such as the anode, cathode, separator, electrolyte, current collector, and casing, become imperative to make equitable comparison possible across different cell formats under suitable approximation, but such information is not always reported.

The performance of a cell depends on materials, synthesis of materials, and cell assembly, so it is imperative to systematically map them. The performance of a system is also strongly influenced by testing conditions such as voltage window, current density, cycle number, temperature, and environmental conditions like humidity, season, etc. Gravimetric charge discharge curves obtained from such different efforts can provide all key performance metrics, including capacity, energy density, power density, coulombic efficiency, and energy efficiency at specific cycles. There are no guidelines for reporting such crucial data across the community, and researchers selectively report only what they wish to highlight. As a result, lots of crucial information goes unreported. On the other hand, reporting datasets of complete charge discharge cycles often lead to excessively large data volumes, so not all data is typically shared. There is a need for a standardized approach to report charge discharge data that preserves critical information while avoiding unnecessary data overload.

There are a few efforts across the community to improve the comparability, standardization, and reliability of performance metrics for these energy storage materials. RG Scurtu et al. [1] highlighted that cell area and format are key to achieving standardization, arguing that largearea wound cells, such as the C21700, deliver more reliable electrochemical data than other smaller-area formats. Similarly, C. Heubner et al. [2] presented a Ragone calculator capable of estimating performance data at the full cell level based on electrochemical measurements on electrodes. Few journals like Journal of Power Sources and Wiley-VCH adopted checklists for batteries, supercapacitors, and related hybrid capacitors to improve the reproducibility and data quality of their published content [3, 4, 5]. Few others have also made efforts to standardize reporting of experimentalists and computationalists by developing checklists [6, 7, 8]. There are a few frameworks/ databases in the community of electrochemical cells that host only a certain class of materials, data from only experiments/ computations, or only a few essential parameters with limited information on dimensions and performance [9, 10, 11, 12, 13, 14].

To address the shortcomings, there is a need for a reporting framework that maps materials, their processing, and cell assembly to cell performance, including gravimetric charge-discharge curves, and makes them comparable. However, to the best of our knowledge, there is no framework that systematically stores all relevant information and maps it to the electrochemical performance. We therefore present a carefully curated, fully selfcontained framework that spans from cell materials, their synthesis and cell assembly, testing conditions, and the physical attributes of cells, such as mass and size, and map them to performance metrics. Using the framework, we developed a simple approach to project cell performance from any small-area laboratory-level cell to any large-area device-level cell using the new specific performance. Leveraging this framework, we have compiled a database (hosted at https://power.tattvasar.com/) by extracting pertinent data from existing literature.

This paper begins by introducing the set of parameters in our framework, designed to enable fair and consistent comparisons. It then discusses the limitations of current reporting practices, particularly on the specific performance normalized by the mass of active material, which do not compare in the same way as specific performance normalized by the total mass of the cell. Next, we present a simple calculator to estimate the specific capacity per unit mass of the total cell of a large-area (C18650) cell using our new mass-normalized specific performance. We explain the competencies of this new specific performance and show its efficient ability to compare the performance of the cell compared to the specific performance normalized by the mass of active material. The paper concludes by presenting the implications of the framework in the research and industrial communities.

## 2 | Detailing the Electrochemical parameters in the Framework

The framework incorporates meticulously curated *parameters* which majorly influence the electrochemical performance of a cell. It ensures reliable and equitable comparability of performance metrics reported across the literature. The parameters are grouped into the following categories to facilitate clear data organization, as described below.

1. **Identification:** The section captures the details of the source of data, such as the Title of the publication, the DOI of the publication, the journal name, and the list of authors.
2. **Materials:** The fields in this section contain information about various material components of a cell, such as anode, cathode, electrolyte, and separator.
3. **Processing:** Details of techniques for the preparation of electrode materials and their assembly procedure into a cell are briefly reported in this section.
4. **Cell Information:** Physical characterization of a cell, such as weights, volumes, and area of different components of a cell, cell type (half or full), and source of analysis, is incorporated in the section.
5. **Testing conditions:** Testing parameters involved in electrochemically characterizing a cell, such as testing voltage window, current density, temperature, and testing environment, are included in this section.
6. **Charge - Discharge profiles and other performance parameters:** Outcomes of the cyclic charge-discharge tests, such as specific capacity, energy density, retention percentage, etc, during electrochemical testing are stored in this section.

The cell performance is highly sensitive to multiple factors, including the choice of materials, their synthesis routes, cell assembly procedures, cell-specific attributes such as component masses and physical dimensions, and electrochemical testing conditions. A framework for systematically mapping these parameters to the cell's performance was developed, as illustrated in Figure 1. $M_i$ indicates the $i^{th}$ combination of materials for cathode, anode, electrolyte, and separator, $S_i$ indicates the $i^{th}$ way of synthesizing a material component of a cell, $CA_i$ indicates the $i^{th}$ procedure of cell assembly, $T_i$ indicates the $i^{th}$ way of testing, $P_i$ indicates the $i^{th}$ resulting performance.
Two different material combinations, $M_1$ and $M_2$, were electrochemically characterized to obtain specific performance. $M_1$ and $M_2$ might differ in the choice of cathode, anode, separator, or electrolyte, either one of them or a few of them, or all of them. The materials in $M_1$ was synthesized using two different techniques $S_1$ and $S_2$. The $S_1$ synthesized $M_1$ material were assembled following $CA_1$ cell assembly procedure. Then, it was tested at three different testing conditions $T_1$, $T_2$, and $T_3$. The obtained performance is mapped in the following way: $M_1 \rightarrow S_1 \rightarrow CA_1 \rightarrow T_1 : P_1$, $M_1 \rightarrow S_1 \rightarrow CA_1 \rightarrow T_2 : P_2$, $M_1 \rightarrow S_1 \rightarrow CA_2 \rightarrow T_3 : P_3$. Another cell assembly procedure $CA_2$ was followed to build a cell for $S_1$ synthesized $M_1$ materials. They were tested at $T_1$ and $T_2$ testing parameters and resulting performance was $P_4$ ($M_1 \rightarrow S_1 \rightarrow CA_2 \rightarrow T_1 : P_4$) and $P_5$ ($M_1 \rightarrow S_1 \rightarrow CA_2 \rightarrow T_2 : P_5$). The $M_1$ materials were synthesized using another technique, such as $S_2$, and two cells were made using two different cell assembly procedures, $CA_1$ and $CA_2$. Both the cells were tested at $T_1$ testing parameters and the resultant performance were $P_6$ ($M_1 \rightarrow S_2 \rightarrow CA_3 \rightarrow T_1 : P_6$) and $P_7$ ($M_1 \rightarrow S_2 \rightarrow CA_4 \rightarrow T_1 : P_7$). Likewise, in material combination $M_2$, the materials are synthesized through $S_1$ and $S_2$ techniques. The $M_2$ materials synthesized through $S_1$ are assembled using two different techniques $CA_1$ and $CA_2$. Each of them are tested at $T_1$ testing parameters to give final performance of $P_8$ ($M_2 \rightarrow S_1 \rightarrow CA_1 \rightarrow T_1 : P_8$) and $P_9$ ($M_2 \rightarrow S_1 \rightarrow CA_2 \rightarrow T_1 : P_9$). Again, the $M_2$ materials synthesized through $S_2$ are assembled using two different techniques $CA_1$ and $CA_2$. Each of them are tested under $T_1$ testing parameters to give final performance of $P_{10}$ ($M_2 \rightarrow S_2 \rightarrow CA_1 \rightarrow T_1 : P_{10}$) and $P_{11}$ ($M_2 \rightarrow S_2 \rightarrow CA_2 \rightarrow T_1 : P_{11}$).

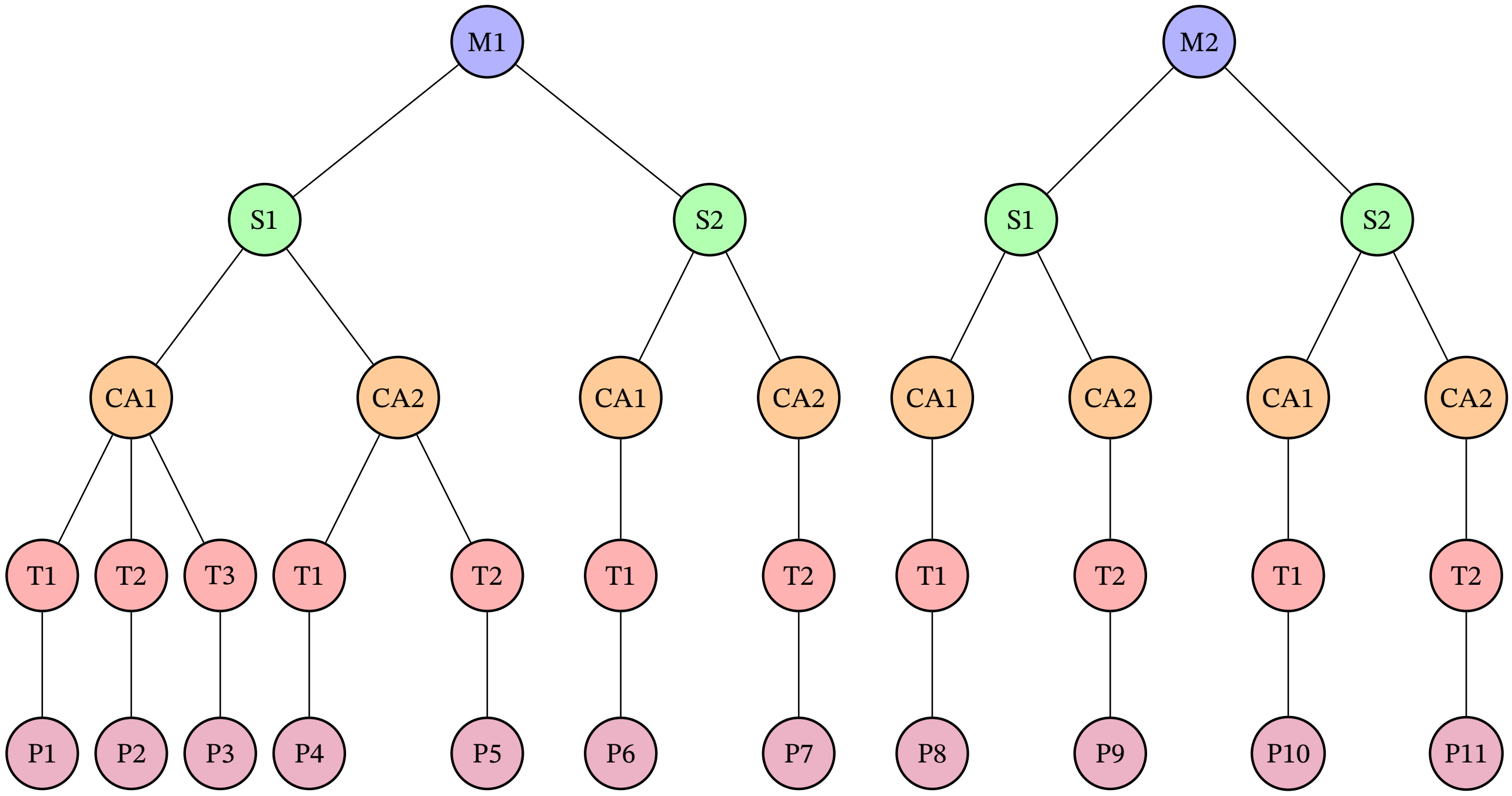

**FIGURE 1** Mapping of materials, their synthesis, cell assembly procedures, and testing to performance. $M_i$ indicates the $i^{th}$ combination of materials for cathode, anode, electrolyte, and separator, $S_i$ indicates the $i^{th}$ way of synthesizing a material component of a cell, $CA_i$ indicates the $i^{th}$ procedure of cell assembly, $T_i$ indicates the $i^{th}$ way of testing, $P_i$ indicates the $i^{th}$ resulting performance.

## 2.1 | Identification

The identification section contains the metadata of a research publication/ articles from which the data is extracted. It can be used to access them when the user is interested in details, which are beyond the scope to be stored in the framework.

The details stored are as follows:

- Name of the publication
- DOI of paper
- Year of publication
- List of Authors
- Name of the journal, archive, patent, private communication

## 2.2 | Materials

- **Shuttling ion:** The ions that travel between electrodes during charging and discharging are called shuttling ions. For example, in lithium-ion batteries, Li is the shuttling ion, and in sodium-ion batteries, Na is the shuttling ion. Recent innovations such as dual-ion batteries, hybrid lithium-sodium ion batteries, mixed-ion batteries cell also exist where multiple ions shuttle between the anode and the cathode [15, 16, 17] can also be stored.
- **Active Cathode and elements involved in Cathode:** We consider the cathode to be that electrode where the reduction reaction takes place during the discharge of the cell. The cathode is composed of active and passive materials, mixed in certain ratios (80:20, 90:10, etc.) to form a slurry that is coated onto the current collector, usually Al. The active material majorly determines the electrochemical capacity of a cell. A slight change in the chemistry of the active material can significantly vary the electrochemical performance of a cell. To capture performance variation, the exact stoichiometry of the active cathode must be reported. Sometimes, the active material is also doped or coated with other elements or materials to enhance the performance of a cell. The elements of the dopants and coatings on the active cathode must also be reported to understand their influence on performance variation.

- **Active Anode and elements involved in Anode:** We consider the anode to be that electrode where the oxidation takes place during the discharge of the cell. Similar to the cathode, the anode is also composed of active and passive materials, mixed in certain ratios (80:20, 90:10, etc.) to form a slurry that is coated onto the current collector, usually Cu. A slight change in the chemistry of active material in the anode can significantly vary the electrochemical performance of a cell, and therefore, the exact stoichiometry of the active anode must be reported. Changes in performance due to dopants or coatings can also be reported to understand their influence on performance variation.
- **Passive materials for cathode and anode:** Passive materials, such as binders and conductive additives, support the active material. The chemistries of passive materials can also vary the overall performance of the cell. They must be reported as mentioned in the publication to understand their influence on performance. E.g., PVDF binder and conductive carbon black at a 13:7 wt% ratio.
- **Electrolyte salt, solvent, and additives:** Electrolyte salt, solvent, and additives play a crucial role in determining the electrochemical performance of a cell. Research is focused on expanding the operating voltage window of the electrolyte by exploring various potential chemistries to replace conventional liquid-based electrolytes [18, 19, 20]. Safety concerns have also pushed the community to work extensively to develop solid-state electrolytes [21]. The chemistries/ brand names of the electrolyte salt, solvent, and additives, exactly as mentioned in the publication, must be reported to capture their influence on the performance. E.g., Gen-2 electrolyte, PEO.
- **Separator:** Similar to electrolyte, the exact chemistries/ brand names of the separator must be reported to capture the change in electrochemical performance of a cell due to them. For capacitors, it is equivalent to a dielectric material, which can be integrated here.

## 2.3 | Synthesis and Cell Assembly

Cell manufacturing can be broadly classified into two stages: *material synthesis* and *cell assembly*.

**Synthesis:** Materials that are involved in various components of a cell could be synthesized in various ways. The influence on performance caused by following different synthesis routes is captured by reporting such information. For example, i) Methods involved in synthesizing an active cathode material, ii) stirring and other parameters to mix active and passive parts to form a cathode slurry, fall under this stage and lead to different performance metrics.

**Cell assembly:** The parameters involved in assembling individual components into a cell must be reported. Parameters such as the ratio of active to passive material in cathode/ anode, electrode thickness coated on current collectors, calendaring pressure applied to make the electrode dense - all of these must be reported.

The primary objective is to capture the variation in the cell's cyclic performance resulting from changes in processes associated with cell manufacturing and assembly, while maintaining the same materials and testing conditions. For example, a change in the pressure of calendaring or a different route for the preparation of electrode material can cause a noticeable change in the specific capacity or energy density [22, 23].

## 2.4 | Cell Information

- **Electrode Area** Energy storage in a cell is directly proportional to the electrode area. Five layers - such as i) anode, ii) cathode, iii) anode and cathode current collectors, iv) electrolyte, and v) separators are the components that proportionately increase as the electrode area of the cell increases. Hence, knowing the performance and total mass of these five components, both per unit area of electrode, is a better metric for comparison across various types of cell formats. Therefore, the area of the electrode becomes a crucial factor, because it helps in translating specific performance interchangeably between cells of different formats, which is explicitly explained in Section 3.1.1.

  → **Total area of the cathode ($cm^2$):** The area of the cathode material coated over Al or positive current collector must be reported.

  → **Total area of the anode ($cm^2$):** The area of the anode material coated over Cu or negative current collector must be reported.

- **Cell component weight considered for specific gravimetric calculations:** Most performance indicatorssuch as specific capacity or energy densityare typically normalized per unit mass. The mass based on which the normalization is done can refer to different component-wise masses - such as active cathode alone, active anode alone, whole cathode (summation of both active and passive parts of the cathode), whole anode (summation of both active and passive parts of the anode), or even the complete mass of the cell. While the research community dealing with small-area cells follows the practice of reporting specific capacity based on active cathode/ anode mass alone, the specifications of commercial large-area cells are generally reported on the basis of total mass. Different component-wise masses are helpful in translating the metrics between normalization standards, i.e., between metrics per active cathode mass to metrics per total cell mass, etc.

- → **Total mass of the active cathode (g):** During cathode fabrication, the active cathode material (bright region) is blended with binders, solvents, additives, and other components (dark region) to form a slurry, which is then coated onto the Al current collector generally (see Figure 2. This includes only the mass of the active part of the electrode, excluding the mass contributed by binders, solvents, additives, and other inactive constituents.
- → **Total mass of the active anode (g):** Similar to the *total mass of the active cathode (g)*, the mass of the active anode, excluding the mass contributed by binders, solvents, additives, coatings, and other inactive constituents, is captured.
- → **Total mass of the cathode (g):** It includes the total mass of the cathode comprising masses of both active and passive constituents (binders, solvents, additives, coatings, and other inactive constituents) of the cathode together, without the mass of its respective current collector.
- → **Total mass of the anode (g):** It contains the total mass of the anode, including both the active and passive parts of the anode, without the mass of its respective current collector.
- → **Total mass of electrochemical participants (g):** It includes the mass of the cathode, anode, metal current collectors, electrolyte, and separators only, which participate in storing the energy, transfer of charges/ ions, etc. These layers proportionately increase as the electrode area of the cell increases. The electrochemically non-participants, like casing, springs, spacers, disc plate, PTC, CID, gasket, rings, etc., are excluded.
- → **Total mass of cell (g):** The total mass of the cell includes the mass of both electrochemical participants and non-participants of a cell.

- **Volume:** The component-wise volume must be reported, as it would help in determining performance metrics normalized per unit volume. Unlike masses, both the experimental and computational groups can report the reliable values of volumes.

- → **Total volume of the active cathode ($cm^3$):** The volume of only the active part of the cathode must be reported.
- → **Total volume of the active anode ($cm^3$):** The volume of only the active part of the anode must be reported.
- → **Total volume of the cathode ($cm^3$):** The total volume of the cathode contains volumes of the active and passive parts of the cathode, as a whole.
- → **Total volume of the anode ($cm^3$):** The total volume of the anode contains volumes of the active and passive parts of the anode as a whole.
- → **Total volume of electrochemical contributors of a cell only ($cm^3$):** It includes the volumes of the cathode, anode, their respective current collectors, electrolyte, and separators, excluding the contribution of volume from the non-contributing electrochemical components like casing, springs, spacers, disc plate, PTC, CID, gasket, rings, etc.
- → **Total volume of cell ($cm^3$):** The cell's total volume contains the volume of both the contributing components and non-contributing components together.

## 2.5 | Testing Conditions

The gravimetric chargedischarge (GCD) test is a cyclic electrochemical characterization performed at a fixed current density over a predetermined voltage window, designed to evaluate a cells performance. The following are the critical testing conditions, which can influence the outcomes of GCD tests.

- **Current Density:** Variations in performance metrics with respect to change in current/ current density can be captured across different materials when explored experimentally. The constant nominal current under which the cell is tested must be reported in milliamperes (mA). However, most publications normalize the current and report it as a current density. The *Current Density* is an amount of applied current normalized mostly by the mass of active material $mAg^{-1}$. With the help of the provided details of component-wise masses, the current density can be translated to any normalization per mass component, such as current density per mass of active cathode alone or current density per mass of total cell. However, the current normalized by mass of active material only compares the performance of a material and not cell. Another way of reporting is C-rate in which, the current density is normalized by the theoretical capacity of an electrode material. As a result, the C-rate becomes material-specific and is not a good option to compare the performance across different cells. It allows for a comparison within the same material only. For example, a 1C rate and a 2C rate of the same material can be compared; however, a 1C rate for one material cannot be compared to 1C rate of another material, making performance comparison across publications difficult. Even within a publication, the C-rate cannot be compared between two different combinations of anode and cathode materials. For example, if a cell made of material A (anode) and B (cathode) is tested at 1C, and the other combination of material A (anode) and material C (cathode) is tested at 1C, the nominal capacity in both combinations does not necessarily mean the same, and hence cannot be compared. Therefore, we strongly recommend that the community report current in mA. For normalized reporting, we urge the researchers to prefer current density in $mAg^{-1}$, a universally adopted unit that enables smooth comparison. Using component-wise masses, the current density normalized by active electrode alone $mAg^{-1}$ can be either converted to nominal current (mA) or translated to current density normalized to any other mass, such as the mass of the total cell. In the case of modeling like DFT calculations, the current density ($mAg^{-1}$) is 0.
- **Temperature:** The temperature corresponding to the electrochemical testing of a cell, either experimentally performed or computationally simulated, must be reported in Kelvin.
- **Testing Environment:** Sometimes, the performance could be even sensitive to the environment in which the testing

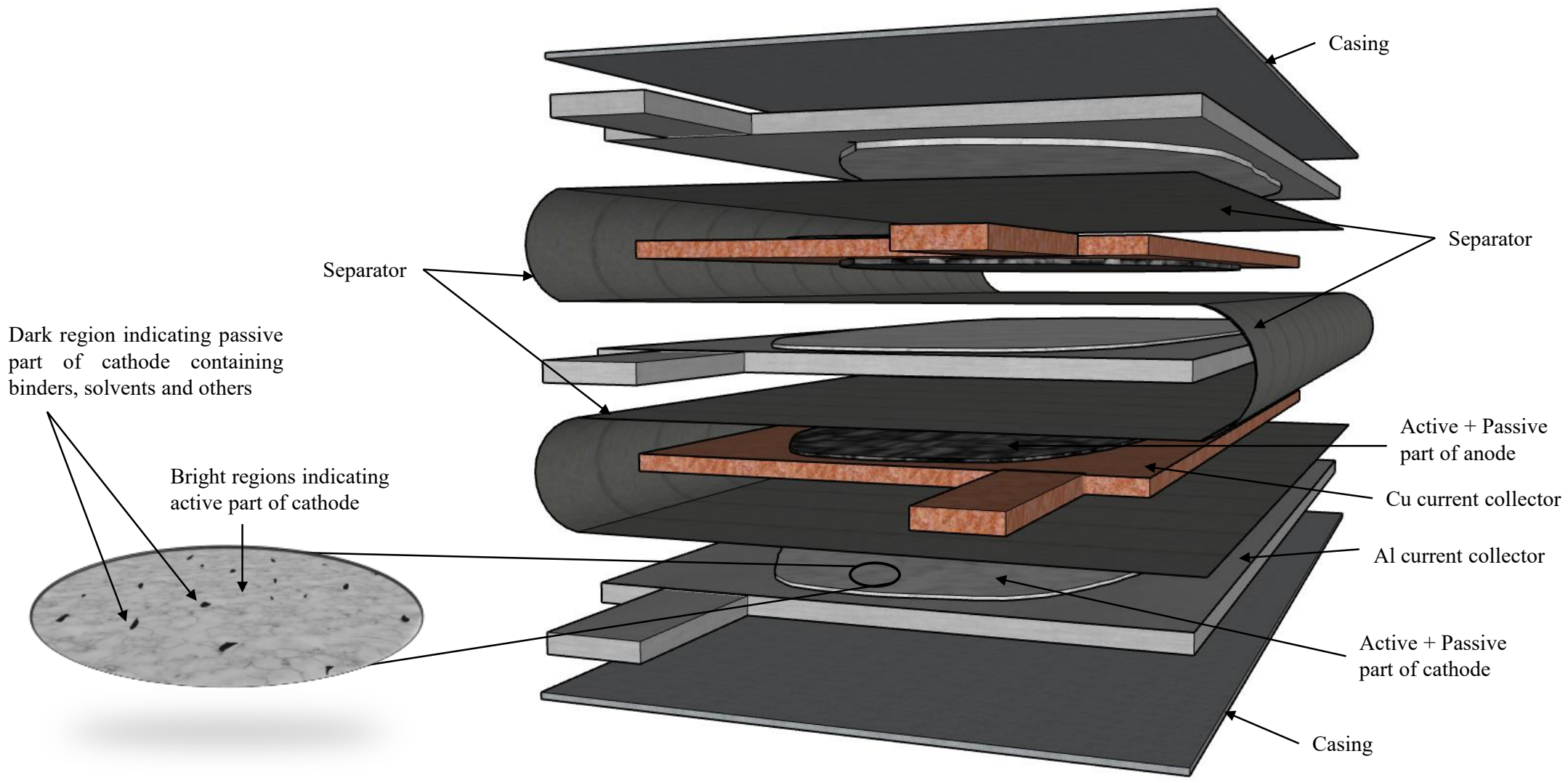

(a) Schematic of Pouch cell (Laboratory-level).

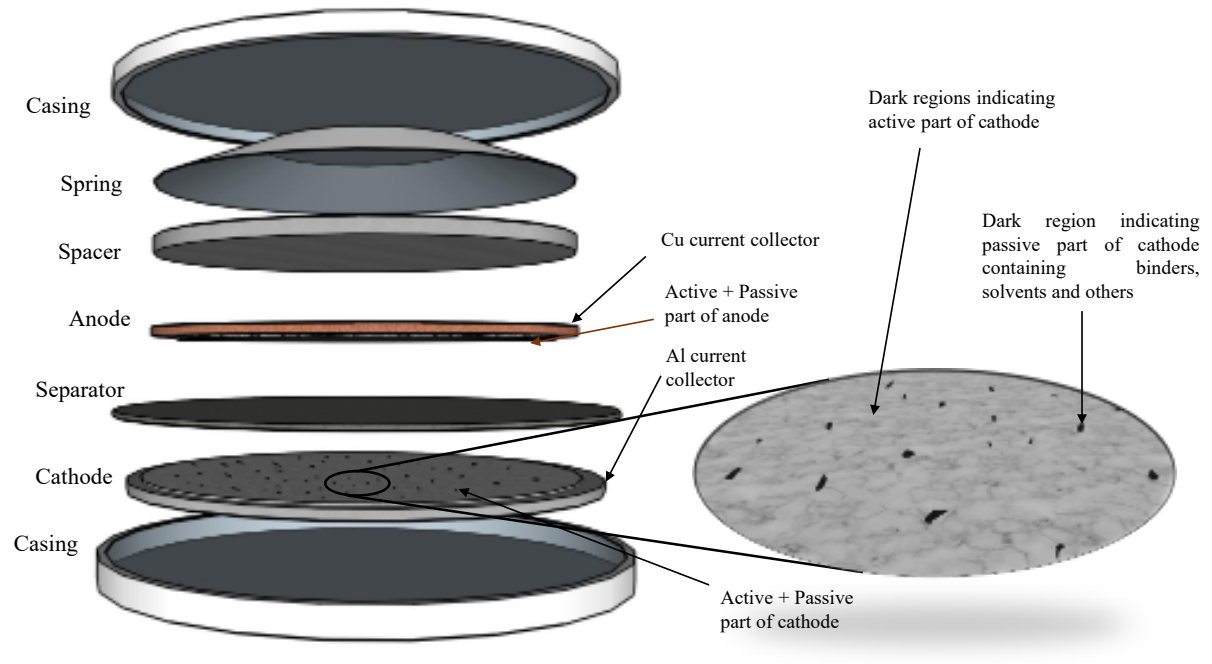

(b) Schematic of Coin cell (Laboratory-level).

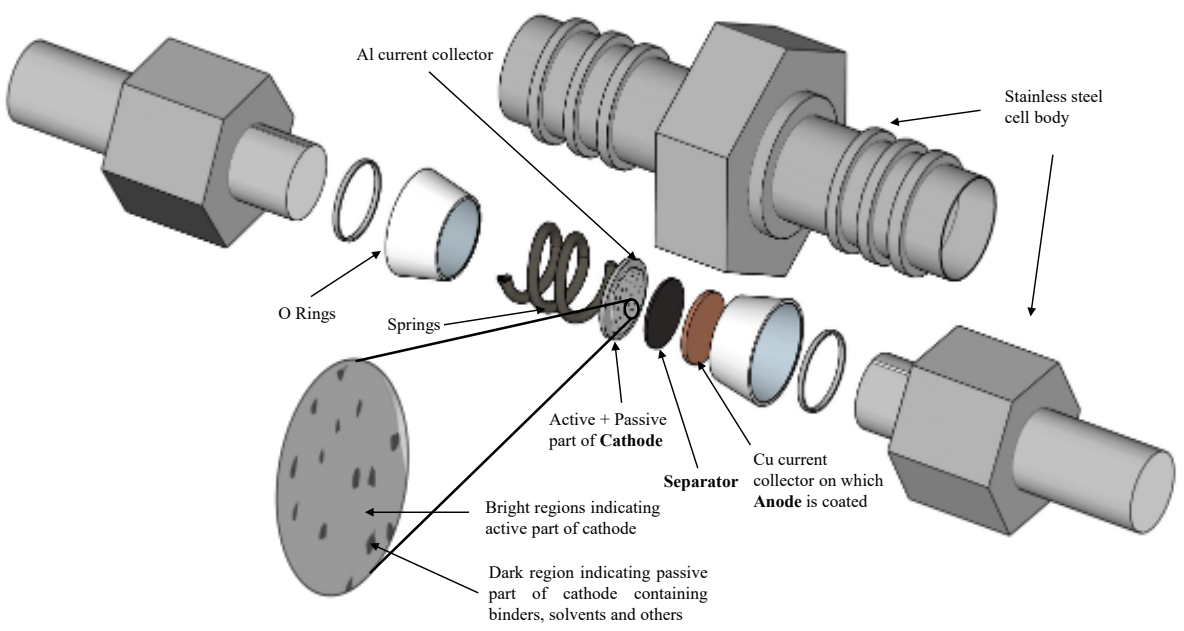

(c) Schematic of Swagelok cell (Laboratory-level).

**FIGURE 2** Schematics of laboratory analysis specimen and device-level cell (Not scaled).

is performed. Humidity on the test day, at the test site, the season, the day, and the time could affect cell performance. The framework even captures how such sensitive parameters can influence the cell performance, which might be critical for cells designed for extreme applications.

- **Minimum and Maximum voltage:** There is a voltage window under which the gravimetric charge-discharge test (GCD) is performed. The minimum and maximum voltage of such tests must be recorded. This information helps in retrieving materials with the highest performance metrics (like discharge specific capacity, or specific energy density, etc.) within the choice of desired voltage window.

## 2.6 | Charge Discharge profiles

The outcome of the GCD test is a voltage as a function of capacity or charge-discharge curves across many cycles. These can also be obtained from computational (DFT) simulations, but only for one cycle. In the case of primary cells, only the first cycle information would be available. These profiles enable the accurate determination of several critical performance metrics, including specific capacity, specific energy density, coulombic efficiency, and energy efficiency. These GCD tests are performed over varying numbers of cycles, like 1000s, 10000s, depending on the capacity retention of the cell. However, when it comes to reporting, they only report data from a few cycles, which they want to highlight. If data of only few cycles are available, say, $1^{st}$, $100^{th}$, $500^{th}$, and $1000^{th}$ cycles, then it becomes challenging for a reader to assess the cells performance at the $300^{th}$ cycle, especially if that aligns with their specific requirements or interests. The interpolation also becomes unreliable because of the limited data points. This happens due to the lack of reporting standards involved in charge - discharge cycles. Conversely, reporting information for all 1000s, 10000s of cycles is also not a viable choice due to the large volume of data. To address the lack of reporting procedure, we have come up with an efficient way of systematically reporting/ storing charge - discharge cycles without significant loss of information. We propose that the charge-discharge cycles must be reported/ stored at regular intervals of capacity retention percentage.
The capacity retention percentage helps in understanding the variation in specific capacity with respect to cycles, and is formulated as shown in Equation 1. The cycle with maximum capacity within the initial few cycles can be considered to be the first cycle capacity for Equation 1, and the capacity retention percentage of this cycle would be 100%. Based on this cycle, the capacity retention percentage of other cycles is determined. It is known that the specific capacity reduces with cycles, and therefore, the capacity retention percentage also decreases with cycles. We propose that the charge-discharge cycles must be reported at regular intervals of capacity retention percentage, such as 90%, 80%, 70%... up to 0% or up to the last cycle (with step-size of 10), or 95%, 90%, 85%... up to 0% or up to the last cycle (with step-size of 5) as shown in Figure 3. The benefit of following such a reporting standard lies in effective summarization without significant loss of information. Test done for either 1000 or 10000 cycles of charge-discharge, can be reported within 20 curves, if the step size of 5 is chosen. Choice of lower step sizes incorporates data from a larger number of cycles, which yields better data sufficiency. Following such a reporting procedure would also enable the collection of sufficient data points for reliable interpolation of performance.

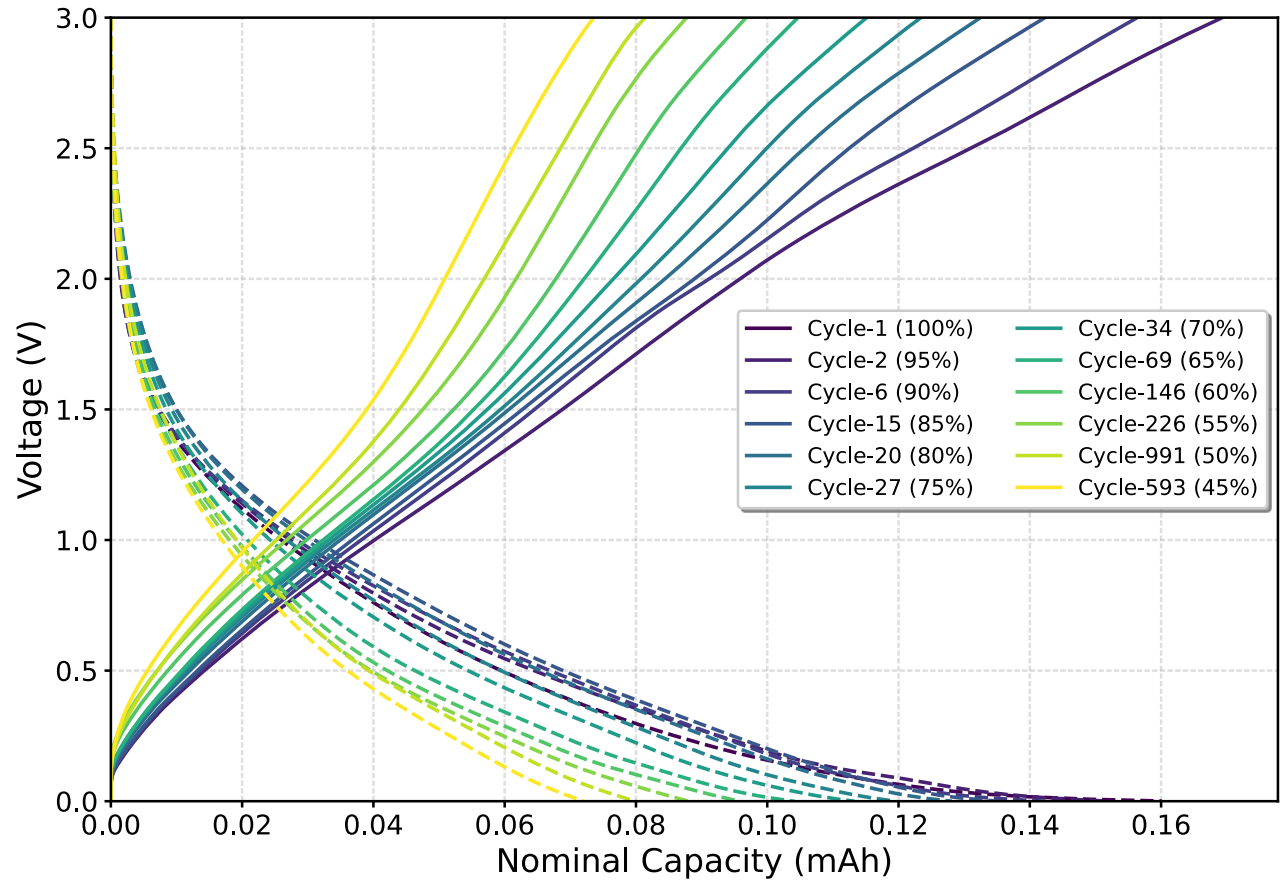

**FIGURE 3** Charge-discharge curves reported at regular intervals of capacity retention percentage at 100%, 95%, 90%, 85%, 80%, 75%, 70%, 65%, 60%, 55%, 50%, and 45%.

$$Retention\ Percentage\ of\ Capacity = \frac{Current\ Cycle\ Capacity}{First\ (Maximum)\ Cycle\ Capacity} \times 100 \tag{1}$$

The charge-discharge curves are self-sufficient to make reliable conclusions about establishing the performance of a cell when GCD tests are performed at different sets of current densities and different sets of voltage ranges for any given cell material. The charge-discharge curves across different voltage ranges would help in figuring out the effective voltage window under which the cell can deliver better performance metrics. The charge-discharge curves along different current densities would help in distinguishing whether the material is suited for energy storage applications or capacitive applications. At lower current density, most of the potential cell materials manifest good energy density. However, higher energy density at higher current density would hint that the material might have better potential in energy storage (battery) applications. Conversely, if energy density is lower but power density is found to be higher, such materials find higher potential in capacitive (capacitor) applications.

For intercalation-type devices, the charge - discharge profile is typically reported as voltage as a function of capacity, whereas for capacitors and supercapacitors, it is commonly reported as voltage as a function of time. Both these reporting conventions are interconvertible. Since the charge-discharge tests are performed at constant current, the constant current can be simply multiplied by time to convert charge/ discharge curves in voltage as a function of time profile to charge/ discharge curves in voltage as a function of capacity profile. For the charging process, the capacity at any instant can be obtained by multiplying the constant current by the corresponding charging time. During discharge, the capacity can be determined by multiplying the constant current by the difference between the instantaneous time and the total charging time. While these reporting methods are interconvertible, the coexistence of different conventions might create unnecessary confusion amongst the novice researchers in the field of energy storage devices.

The following information can be reliably extracted from GCD curves, which are crucial in assessing the cyclic performance of a cell, provided the cycle number for each curve must be reported.

- **Discharge/ Charge Capacity ($mAh$):** One of the crucial metrics to rank the material performance is capacity. The discharge/ charge capacity is the total amount of electric charge delivered/ stored by a cell or electrode during discharging/ charging. The discharge or charge capacity can be directly obtained from charge - discharge cycles, because they are reported as voltage as a function of capacity.
- **Retention Percentage of Capacity:** The capacity retention percentage between cycles can be obtained from Equation 1.
- **Discharge/ Charge Energy ($Wh$):** The discharge/ charge energy measures the amount of energy that an energy storage device can deliver or store. It is an important metric for energy storage devices, especially for batteries. Many experimentalists report energy using Equation 2.

$$Energy = Capacity\ (Q) \times Voltage\ (V) \tag{2}$$

Although this method is widely used for calculating specific energy in the electrochemical community, it is worth noting that it is an approximation and does not yield an exact value. The voltage keeps changing during charge/discharge in GCD tests. However, a single value of voltage (approximately the midpoint of the operating voltage window) is introduced in Equation 2. The correct way to calculate energy for a particular cycle is to integrate voltage with respect to capacity as shown in Equation 3, yielding the area under the charge-discharge curve for that particular cycle [24]. Both the gravimetric discharge/ charge energy can be obtained from the method of integration and is more accurate.

$$Energy = \int_0^{Q_{max}} V dQ \tag{3}$$

where Q is the capacity, $Q_{max}$ is the maximum capacity, and V is the voltage.

- **Power ($W$):** The power is defined as the rate at which the device delivers or absorbs electrical energy. It basically describes how fast a stored electrochemical energy can be delivered to an external load. Most experimental methods approximate the voltage to a single value and calculate the power using Equation 4.

$$Power = \frac{Q \times V}{t_{d/c}} \quad (4)$$

where Q is the capacity, V is the voltage, and $t_{d/c}$ is the time of discharge/ charge.

The most accurate value of power is achieved through Equation 5, which includes the area under the charge-discharge curves obtained by the method of integration.

$$Power = \frac{\int_0^{Q_{max}} VdQ}{t_{d/c}} \quad (5)$$

where Q is the capacity, $Q_{max}$ is the maximum capacity, V is the voltage, and $t_{d/c}$ is the time of discharge/ charge.

- **Capacitance ($F$):** Capacitance is an important parameter to evaluate the performance of the capacitive behavior of energy devices. It measures how effectively a material stores electrostatic charge. It is defined as the charge stored per voltage change. Capacitance is estimated using Equation 6, but this equation cannot yield an accurate value because the voltage is not constant.

$$Capacitance(C) = \frac{I \times t_{d/c}}{\Delta V} \quad (6)$$

where I is the current, $t_{d/c}$ is the time of discharge/ charge, and V is the voltage.

The right way to calculate capacitance is again using the area under the charge - discharge curves obtained by the method of integration as shown in Equation 7.

$$Capacitance(C) = \frac{\int_0^{Q_{max}} VdQ}{\Delta V^2} \quad (7)$$

where Q is the capacity, $Q_{max}$ is the maximum capacity, and V is the voltage.

- **Discharge Areal Specific Capacitance ($Fcm^{-2}$):** The Discharge Areal Specific Capacitance is a metric used mainly for supercapacitors, which measures how much capacitance is obtained per unit electrode area during discharge.
- **Energy efficiency:** The energy efficiency measures how much energy is conserved during a charge - discharge cycle. Equation 8 is used to obtain energy efficiency. The charge and discharge energies can be obtained from charge-discharge curves using the method of integration, as explained above. The energy efficiency can also indirectly infer the internal resistance of a cell, which is caused by electrolyte ionic resistance, electronic resistance of electrodes, current collector resistance, contact resistance between particles, etc.

$$Energy\ Efficiency = \frac{Discharge\ Energy}{Charge\ Energy} \times 100 \quad (8)$$

- **Coulombic efficiency:** The coulombic efficiency is the ratio of the discharge gravimetric capacity/ specific capacity and charge gravimetric capacity/ specific capacity. It measures the effectiveness with which the charge is transferred during the charge and discharge cycles. It accounts for irreversible reactions like SEI formation and other parasitic redox reactions, which reflect the self-discharge behavior of a cell. Equation 9 is used to obtain coulombic efficiency, where the capacity values are obtained from charge-discharge curves.

$$Coulombic\ Efficiency = \frac{Discharge\ Capacity}{Charge\ Capacity} \times 100 \quad (9)$$

## 2.7 | Few Other important performance parameters

**Leakage current:** Leakage current profiles provide important insights into the self-discharge behavior and interfacial stability of supercapacitor electrodes. By monitoring the decay of current under a constant voltage over time, valuable insights on self-discharge behavior, charge redistribution, ion diffusion, parasitic reactions, electrolyte stability, and internal resistive losses can be obtained, making it an important parameter for assessing the reliability and long-term performance of supercapacitor devices.
**Active material volume expansion:** The change in the volume of the active cathode or active anode during testing in terms of percentage change can provide essential insight regarding electrode stability over many cycles. Although such minute dimensional variations are difficult to quantify experimentally, computational methods such as DFT can capture them accurately.

## 2.8 | Source type Identifiers

Attempts to discover new materials are not just limited to experimental groups, but also the computational-driven research groups (like first-principles DFT calculations), and machine-learned discovery has also grown in recent years. Therefore, it is important to indicate the type of analysis/ approach that resulted in the origination of collected data - whether it is experimental, computational, machine learning, or others. This is because the results from practical experiments are always considered to be ground-truth, whereas other approaches have assumptions, and, therefore, their boundary of reliability needs to be understood before making claims. Therefore, predictions from such approaches can serve as a useful guide for experimentalists to choose from a vast search grid and further understand the practical performance along with the feasibility of device fabrication and assembly.
Computational analysis, like DFT calculations, most often models a single component of a cell individually and understands its theoretical capacity or conductivity of shuttling ions, etc. The reported performance, however, would inevitably vary at the cell level when an individual component is assembled with other materials to form a cell. It must also be noted that the computational models are limited to be assessed at 0 current density and are studied only for one and first cycle, whereas experiments, in contrast, extend their analysis for many cycles. For example, few computationalists compute specific capacity at multiple ion concentrations. Such studies report the voltage (Open Circuit Voltage, OCV) as a function of capacity. Upon interpolating all the points, the curve would resemble an experimentally derived gravimetric charge-discharge curve, but only for one cycle, and can be accommodated in our framework as a charge-discharge curve. However, it is clearly specified that such curves are computationally generated.

Results from machine-learning-based techniques, especially when exploring different material chemistries, are highly sensitive to the type of data used to train the models. Such trained machine-learned models using experimental and computational data predict various performance metrics, charge-discharge curves, etc. However, all the machine-generated data indicates that the data originated using ML-based approaches.

## 3 | Need for a new mass normalization method to report metrics

The best way to compare the efficacy of materials, synthesis, and assembly techniques is the specific performance normalized by mass of a cell (will refer to it as $SP_{TM}$) of the fixed cell size/ format. However, not all exploratory efforts are conducted within a single cell format. In fact, most of them are carried on various small-area cells such as coin cell, swagelok cell, and pouch cell, etc., and the specific performance is reported by normalizing it by the mass of active cathode alone (will refer to it as $SP_{AM}$). The $SP_{AM}$ is a good metric to assess the potential of a material as an active electrode. It is also a good way of reporting to compare the results of computational simulations. However, it does not help in comparing the performance of a cell in the way, as done by specific performance normalized by the total mass of a cell ($SP_{TM}$). It is because the active material alone does not participate in electrochemical energy storage. To form a complete cell device capable of storing energy, the active material must be integrated with the passive material, electrolyte, separator, and current collectors. Therefore, the mass of all of them must be taken into account along with the mass of active material to understand the efficacy of a cell. Moreover, the amount of passive material and the electrolyte depend strongly on the type of active material. For example, the ratio of active to passive materials is generally above 80:20 for most of the electrodes made of bulk materials. In the case of emerging 2D materials which show promising results due to their large number of electrochemical sites per unit mass and fast kinetics associated with the intercalation and de-intercalation of shuttling ions. Conversely, these 2D materials may require different ratios of active to passive components (60:40) compared to bulk materials to integrate as a cell [25]. Hence, a high value of $SP_{AM}$ in the case of 2D materials does not guarantee better performance of a cell. To summarize, the $SP_{AM}$ does not help in comparing the performance of a cell across different cell formats, because it does not account for other electrochemical participants other than the active material itself. Therefore, a new mass normalization method to report specific performance is required, which can account for all the electrochemical participants and compare the performance of a cell across various cell formats.

The $SP_{TM}$ of a fixed large area cell can be estimated from measurements made at a small area cell. The area of the electrode increases from a small-area cell to a large-area cell. It is fair to assume that the mass of electrodes, their current collectors, electrolyte, and separators, which participate in the electrochemical reactions associated with energy storage, also increases in the same proportion as the area of the electrode increases from a small-area cell to a large-area cell. Hence, we propose to define a new mass normalized factor to report specific performance $SP_{PM}$: specific performance normalized to the sum of the mass of the electrodes, their current collectors, electrolyte, and separators as shown in Equation 10. We call this summation of the mass to be the mass of electrochemical participants.

$$SP_{PM} = \frac{\text{Performance } \in \text{ (Capacity, Energy, \dots)}}{M_{\text{anode}} + M_{\text{cathode}} + M_{\text{current collectors}} + M_{\text{electrolyte}} + M_{\text{separator}}} \tag{10}$$

Where, performance can be capacity, energy, etc., $M_{anode}$ is the mass of anode, $M_{cathode}$ is the mass of cathode, $M_{current\ collector}$ is the mass of current collectors, $M_{electrolyte}$ is the mass of electrolyte, and $M_{separators}$ is the mass of separators.

This new specific performance has the ability to compare the efficacy of materials, processing, and cell assembly across various cell formats and the thickness of the active and passive parts of the electrode, which is explicitly explained using the illustration below.

We consider an example of a 2032-type coin cell with carbon-coated $Na_3V_2(PO_4)_3$ (C-NVP) as a cathode and activated carbon as the counter electrode [26, 27]. This coin cell is labeled coin cell-A in Figure 4. The carbon-coated $Na_3V_2(PO_4)_3$ active cathode material of 0.0067 g was mixed with passive material in the ratio of 80:20 to produce a final cathode mixture of 0.0084 g, which was pasted on an Al foil of 15 $\mu m$. The activated carbon of 0.0042 g was mixed with passive material in the ratio of 80:20 to make an anode slurry of 0.0053 g, which was coated on Cu foil of 15 $\mu m$. The mass of the coin cell with electrochemical participants is 0.086 g. The total mass of the cell, including both electrochemical participants and non-participants, is 3.5779 g. A uniform loading mass of 3.81 $mg/cm^2$ and 2.38 $mg/cm^2$ for cathode and anode were considered, respectively. The detailed component-wise masses of coin cell-A are tabulated in supporting information. The area of cathode and anode in coin cell-A is 1.7663 $cm^2$ each. A specific capacitance of 62 $Fg^{-1}$ (based on two electrode masses) was reported for this asymmetric cell, when tested at 0.1 $Ag^{-1}$ in the voltage range of 0.0 - 3.0 V. The specific capacitance is converted to specific capacity $mAhg^{-1}$ using the Equation 11. A $SC_{AM}$ of ≈ 52 $mAhg^{-1}$ is obtained when the specific capacity is normalized by mass of both the electrodes, and 104.47 $mAhg^{-1}$ is obtained when the specific capacity is normalized by mass of active cathode alone. Hence, the cell delivered a nominal capacity of 0.7 mAh (104.39 × 0.0067).

$$Q(mAhg^{-1}) = \frac{C_s(Fg^{-1}) \times \Delta V}{3.6} = \frac{62 \times 3}{3.6} \approx 52 \tag{11}$$

where Q is specific capacity, $C_s$ is the specific capacitance, and $\Delta V$ is the difference between maximum and minimum voltage.

### 3.1 | Estimating performance of large-area cell from small-area cell

In this discussion, we will explain the steps to estimate the $SC_{TM}$ of a large-area C18650 cell from the reported nominal capacity of a small-area coin cell, with the help of an illustration. We will show how our newly proposed mass normalization method helps in estimating cell performance at any format and makes it comparable across different cell formats. Hereon, the capacity normalized per mass of active material, capacity normalized per mass of total cell, and capacity normalized per mass of electrochemical participants only (newly proposed mass normalization method to report specific performance) are referred to as $SC_{AM}$, $SC_{TM}$, and $SC_{PM}$. It must be noted that the area and non-participating mass of the C18650 cell are considered to be 700 $cm^2$ and 10.45 g, respectively, for this illustration [28].

#### 3.1.1 | Steps in the method of estimation

The method to estimate the $SC_{TM}$ of a large-area C18650 cell from the reported nominal capacity of a small-area coin cell-A is discussed below with the help of an illustration.

1. **Computing the scaling factor:**
The scaling factor is the ratio of the area of the cathode (active and passive part of the cathode) in cell format, which is chosen to estimate (here, a large area C18650 cell) and the cell format from which you want to estimate from (here, small area coin cell-A), as expressed in Equation 12.

$$Scaling\ factor = \frac{A_{cell\ format\ of\ choice}}{A_{cell\ format\ reported}} \qquad (12)$$

where $A_{cell\ format\ of\ choice}$ is the area of the cathode in cell format which is chosen to estimate, and $A_{cell\ format\ reported}$ is the area of the cathode in reported cell format from which you want to estimate.
The area of the cathode and anode of small-area coin cell-A in the illustration is 1.7663 $cm^2$. The typical area of a cathode and anode of a large-area C18650 cell is considered to be 700 $cm^2$. Therefore, the area of a small-area cell and a large-area cell are 1.7663 and 700 $cm^2$, respectively, and the resulting scaling factor comes out to be 396.32 ($\frac{700}{1.7663}$).
2. **The mass of electrochemical participants of a small-area coin cell**
The mass of electrochemical participants includes the mass of electrodes, current collectors, electrolyte, and separators. In the depicted illustration, the mass of electrochemical participants of the coin cell-A is 0.086 g.
3. **Calculating $SC_{PM}$:**
The $SC_{PM}$ is the specific capacity normalized by the mass of electrochemical participants. For this illustration, the nominal capacity and mass of the electrochemical participants of coin cell-A are 0.7 mAh and 0.086 g, respectively. Therefore, the $SC_{PM}$ is found to be 8.14 $mAhg^{-1}$ ($\frac{0.7}{0.086}$).
4. **Estimating the mass of electrochemical participants of large-area C18650 cell:**
The mass of electrochemical participants of a small-area cell is multiplied by a scaling factor to translate it to the mass of electrochemical participants of a large-area C18650 cell. Therefore, in this illustration, the mass of electrochemical participants translated to the C18650 large-area cell is 34.08 g (0.086 $\times$ 396.32).
5. **Estimating the total mass of large-area C18650 cell:**
The mass of non-electrochemical participants includes casing, spacer, springs, etc. For this illustration, it is assumed to be 10.45 g for a C18650 cell. This mass is then added to the estimated mass of electrochemical participants to estimate the total mass of a large-area C18650 cell. Therefore, the estimated total mass of the C18650 cell is 44.53 g (34.08 + 10.45).
6. **Estimating the nominal capacity at large-area cell:**
The nominal capacity of a large-area C18650 cell is estimated by multiplying the nominal capacity of coin cell-A by the scaling factor. Therefore, the estimated nominal capacity at a large-area C18650 cell is 277.42 mAh (0.7 $\times$ 396.32).
7. **Estimating the $SC_{TM}$ of C18650 large-area cell:**
The estimated nominal capacity of a large-area C18650 cell is divided by the estimated total mass of the C18650 cell to give an estimated $SC_{TM}$ of 6.23 $mAhg^{-1}$ at a large-area C18650 cell.

The steps of estimation explained above indicate that the mass of electrochemical participants of a cell plays a vital role in translating performance metrics across different cell formats. The $SC_{PM}$ would be 8.14 $mAhg^{-1}$ for both small-area coin cell-A (0.7/0.089) and large-area C18650 cell ($\frac{0.7\times396.32}{0.089\times396.32}$). It must also be noted that $SC_{AM}$ remains the same both for the small-area coin cell and the large-area C18650 cell. For example, the $SC_{AM}$ of coin cell-A is considered to be 104.47 $mAhg^{-1}$ ($\frac{0.7}{0.0067}$) in the illustration, because the nominal capacity and mass of active cathode material are considered to be 0.7 $mAh$ and 0.0067 g, respectively. The estimated $SC_{AM}$ of a large-area C18650 cell will also be 104.47 $mAhg^{-1}$ ($\frac{0.7\times396.32}{0.0067\times396.32}$), because the scaling factor in the illustration computes to be 396.32. Both $SC_{AM}$ and $SC_{PM}$ are the same for both the small-area coin cell and the large-area C18650 cell, thereby remaining independent of the cell format. Therefore, the specific performance normalized by the mass of electrochemical participants $SP_{PM}$ enables a fair comparison of the efficacy of materials, synthesis, and their cell assembly across various cell formats. This new mass normalization approach can also be used to optimize the thickness of the active and passive parts of the electrode, as explicitly explained in the following section.

#### 3.1.2 | Efficacy of newly proposed mass normalization method in optimizing thickness

Evidence in the literature shows that thick coating leads to inefficient utilization of active materials and hence the $SC_{AM}$ decreases with increase in coating thickness of electrode material on a current collector, especially at higher current rates [29, 30, 31, 32] references. The thickening of the cathode reduces the $SC_{AM}$ because the active ions lying in deeper layers experience high ionic transport resistance due to long ionic pathways extending from cathode to anode, decreasing the overall nominal capacity [33]. In order to illustrate the decreasing trend of specific capacity with an increase in thickness of the active and passive part of the electrode material coated over the current collector, a total of 8 coin cells were considered. As explained earlier, the coin cell-A in Figure 4 is the only coin cell that is experimentally determined. The $SC_{AM}$ for the remaining seven coincells (with different thicknesses of active and passive parts of the electrode material) was assumed on the basis of the measured $SC_{AM}$ of coin cell-A to represent the decreasing trend of $SC_{AM}$ with respect to electrode thickness. Their component-wise masses were also carefully scaled on the basis of the component masses of coin cell-A to reflect the prescribed variations of electrode thickness. In all eight coin cells, it was assumed that there would be no change in the mass of the separator, electrolyte, or electrochemically non-participants. The detailed information of assumed capacities and component masses of all 8 coin cells is given in the supporting information.

Figure 4(a) depicts a descending trend of $SC_{AM}$ of small-area coin cells from coin cell A to H. For instance, coin cell-A exhibited a nominal capacity of 0.7 mAh; it would be anticipated that the nominal capacity would double to 1.4 mAh when the thickness is doubled, yet the nominal capacity of coin cell-B has a nominal capacity of 1.35 mAh. Similarly, the nominal capacity anticipated for an increase in thickness by 3 times (coin cell-C), 3.5 times (coin cell-D), 4 times (coin cell-E), 4.5 times (coin cell-F), 5 times (coin cell-G), and 5.5 times (coin cell-H) was expected to be 2.1 mAh, 2.45 mAh, 2.8 mAh, 3.15 mAh, 3.5 mAh, and 3.85 mAh, respectively. As cathode thickness increases, active material located in deeper layers near the current collector experiences greater ionic transport limitations due to longer diffusion pathways. Consequently, these regions become progressively underutilized, effectively behaving as inactive material during charge and discharge, especially at higher current rates. Therefore, the actual nominal capacities come out to be 1.96 mAh, 2.22 mAh, 2.38 mAh, 2.4 mAh, 2.42 mAh, and 2.42 mAh, respectively. Since the $SC_{AM}$ is the ratio of nominal capacity and mass of the active cathode alone, although both the numerator and denominator keep increasing, the numerator does not increase with the same factor as the denominator. This causes $SC_{AM}$ to decrease with increasing electrode thickness. Coin cell-A is found to have the optimal (highest) $SC_{AM}$ among all the coin cells. The value of $SC_{AM}$ remains the same for both the small-area coin cell and the large-area C18650 cell. Figure 4 shows $SC_{AM}$ for both small-area coin cell and large-area C18650 cell, showing the decreasing trend with increase in electrode thickness. We now utilize the nominal capacity of all coin cells A to H to estimate the corresponding $SC_{PM}$, and $SC_{TM}$ of a small-area 2032-type coin cell and a large-area C18650 cell, and examine their behavior with the change in electrode thickness.

The same methodology explained in Section 3.1.1 is applied to all the coin cells B to H to estimate the $SC_{PM}$ and $SC_{TM}$ of small-area 2032-type coin cells and large-area C18650 cells from the reported values of nominal capacity of small-area coin cells. In all cases, the area of the electrode and the non-participating mass of the C18650 cell were considered to be 700 $cm^2$ and 10.45 g, respectively. For this illustration, the same scaling factor of 396.32 is utilized for the estimation of specific capacity and masses of large-area C18650 cells from the small-area coin cells. To estimate $SC_{TM}$ of a small-area coin cell, the nominal capacity of the coin cell is divided by the total mass of the coin cell obtained by adding the masses of electrochemical participants and non-participants. For this illustration, the area of the electrode and the non-participating mass of the small-area coin cell were considered to be 1.7663 $cm^2$ and 3.49 g, respectively. For coin cell-A, the $SC_{TM}$ of the small-area coin cell comes out to be 0.20 $mAhg^{-1}$ ($\frac{0.7}{0.086+3.49}$). The $SC_{TM}$ of small-area coin cells for all other coin cells B to H are calculated using a similar methodology.

Figure 4(a) shows the estimated $SC_{TM}$ of small-area coin cell and large-area C18650 cell from nominal capacity of all coin cells A

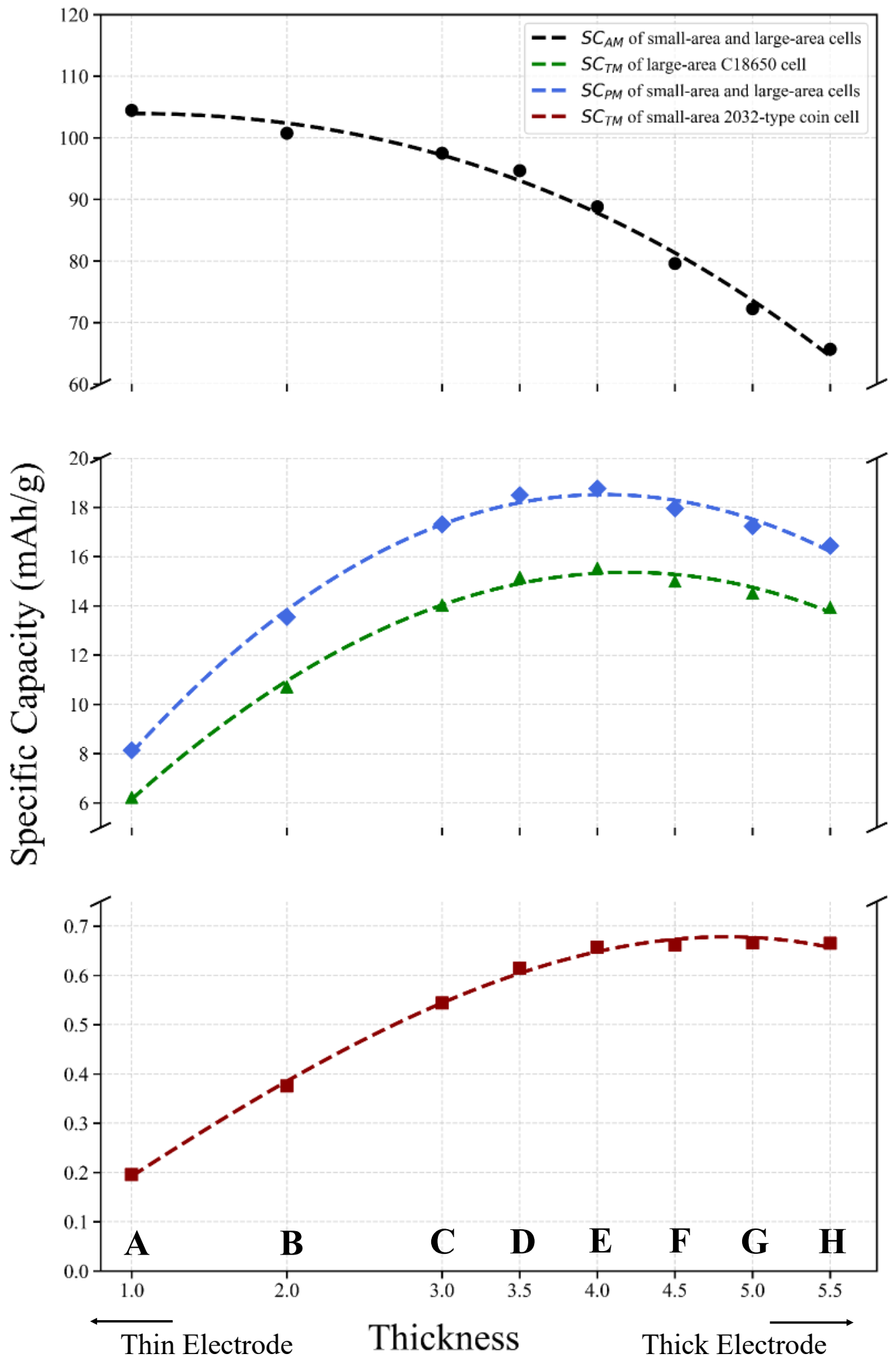

(a) Black line indicates the change in specific capacity normalized per mass of active cathode ($SC_{AM}$) of small-area 2032-type coin cell and large-area C18650 cell with change in electrode thickness. Blue line indicates the change in specific capacity normalized per mass of electrochemical participants ($SC_{PM}$) of small-area 2032-type coin cell and large-area C18650 cell with change in electrode thickness. Green line indicates the change in specific capacity normalized per total mass of cell ($SC_{TM}$) of large-area C18650 cell with change in electrode thickness. Dark red line indicates the change in specific capacity normalized per total mass of cell ($SC_{TM}$) of small-area 2032-type coin cell with change in electrode thickness.

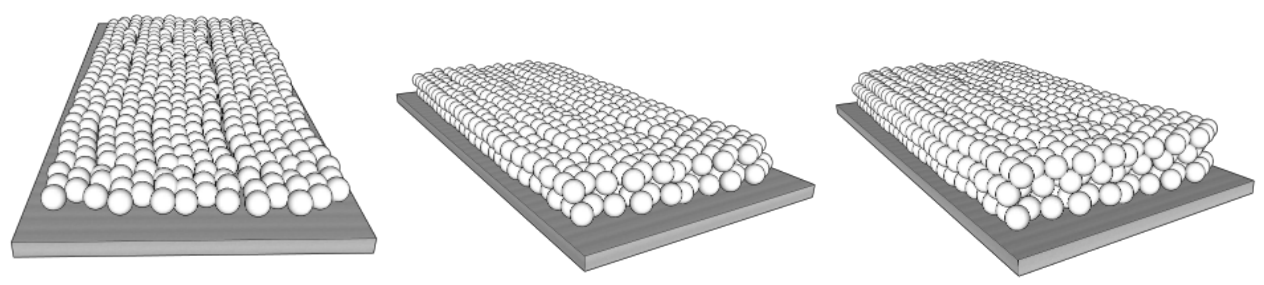

(b) Coin cell-A (c) Coin cell-B (d) Coin cell-C

**FIGURE 4** Change in specific capacity of small-area and large-area cells with respect to active cathode thickness.

to H. Unlike the trend of $SC_{AM}$, the estimated $SC_{TM}$ of a small-area and large-area cell increases steadily up to the thickness of coincell-E, beyond which it plateaus and becomes insensitive to further thickening of the cathode. In other words, once the cathode thickness exceeds the coincellE thickness, the active material located in deeper layers near the current collector no longer contributes to storage capacity, effectively becoming inactive or dead material. Therefore, the trends of $SC_{AM}$ and $SC_{TM}$ are different both for small-area and large-area cells. Researchers working with any cell format often coat thin layers of electrode (active and passive parts) to maximize the $SC_{AM}$, as shown by the $SC_{AM}$ of coin cell-A. Based on these inflated values of $SC_{AM}$, they make strong claims about the performance. It must be noted that the performance reported in terms of $SC_{AM}$ corresponds to the material and not the cell. The $SC_{TM}$, which reflects the performance of a cell, does not maximize at coin cell-A with thin electrode coating as shown in Figure 4. Instead, the optimal (highest) performance of the cell is found at coin cell-E with a thicker coating than coin cell-A, because the $SC_{TM}$ of the small-area coin cell and the large-area C18650 cell maximizes at coin cell-E. Though a similar rise and saturation pattern is evident in both $SC_{TM}$ of small-area and large-area cells, the magnitude of $SC_{TM}$ differs between small-area and large-area cells. Therefore, unlike $SC_{AM}$, the $SC_{TM}$ compares the performance of a cell. However, the limitation of $SC_{TM}$ is that the magnitudes of $SC_{TM}$ change with the cell formats, even though the trends are preserved across cell formats. This is because the ratio of active mass to the rest of the mass differs across different cell formats, which, in turn, makes the comparison difficult. To summarize, i) even though $SC_{AM}$ is independent of cell format, it does not compare the performance of a cell in a way, as done by $SC_{TM}$, rather it corresponds to performance of a material only ii) though the trends of $SC_{TM}$ remains same across the cell formats, their magnitudes changes, making it dependent on cell format and difficult to compare.

We now see how our newly proposed mass normalization method for reporting specific capacity $SC_{PM}$ overcomes the shortcomings of $SC_{AM}$ and $SC_{TM}$. Figure 4(a) shows the behavior of $SC_{PM}$ with a change in electrode thickness across cells A to H. It is observed that the $SC_{PM}$ depicts the trend of $SC_{TM}$ of the small-area coin cell and large-area C18650 cell. It is also found that the $SC_{PM}$ also optimizes the thickness of the electrode at coin cell-E, as done by $SC_{TM}$. Moreover, the $SC_{PM}$ is the same for both the small-area coin cell and large-area C18650 cell, indicating that $SC_{PM}$ remains consistent across different cell formats, making it more comparable. Therefore, the $SC_{PM}$ is a unique number that is independent of cell formats, electrode thickness, and captures the performance of a cell, effectively addressing the limitations associated with both $SC_{AM}$ and $SC_{TM}$.

Our newly proposed mass normalization method to report any performance metrics, $SP_{PM}$, therefore has the following competencies:

1. It is the same across different cell formats for chosen material, synthesis, and cell assembly. This enables a fair comparison of the efficacy of materials, synthesis strategies, and cell assembly techniques across cells of different formats.
2. It compares the performance of a cell in a way, as done by specific performance normalized by the total mass of a cell, which is mostly preferred.
3. It is also useful in translating/ estimating $SP_{TM}$ of any cell formats from the reported values of $SP_{PM}$ of any cell format.
4. It can be used to optimize the thickness of the electrode in terms of cell performance, like specific capacity, energy density, etc. It must be noted that the optimum thickness may be different for different specific performance metrics. For example, in the above example, the optimum thickness for specific capacity was found at coin cell-E. The optimum thickness for energy density may be found in different coin cells with different thicknesses. Optimal loading of active material in the cell can also be estimated. Cell performance per unit price can also be calculated. These information helps cell manufacturers and industries to test and commercialize the laboratory-explored cells into consumer-ready device-level cells.

### 3.1.3 | Cell performance from the perspective of reporting standards for Current

Till now, we explained how reporting $SP_{AM}$ can mislead in terms of performance of a cell. In this section, we will show, how a coin cell with thicker electrodes like coin cell-E (4 times higher compared to that of coin cell-A) is favored in terms of cell performance from the perspective of reporting standards for current. For illustration, the coin cell-A and coin cell-E from the previous illustration is considered. $SC_{PM}$ and $SC_{TM}$ indicate that coin cell-E performs superior over coin cell-A, when $SC_{AM}$ shows that coin cell-A performs superior to coin cell-A. The theoretical specific capacity of $Na_3V_2(PO_4)_3$ is considered to be 117.60 $mAh$. For this illustration, the theoretical specific capacity of $Na_3V_2(PO_4)_3$ is considered to be theoretical specific capacity of a cell. The area of the cathode of coin cell-A and coin cell-E are the same (1.7663 $cm^2$). The mass of the active cathode of coin cell-A and coin cell-E are 0.0067 g and 0.0268 g, respectively. If both the coin cells are subjected to electrochemical testing by drawing current of 0.67 mA, the areal current would be the same for both the coin cells, 0.3793 $mAcm^{-2}$ ($\frac{0.67}{1.7663}$). The current density normalized to mass of active cathode alone for coin cell-A and coin cell-E would be 100 $mAg^{-1}$ ($\frac{0.67}{0.0067}$) and 25 $mAg^{-1}$ ($\frac{0.67}{0.0268}$), respectively. However, the C-rate of coin cell-A would be 0.85C ($\frac{100}{117.60}$), which is four times higher than that of coin cell-E, 0.21C ($\frac{25}{117.60}$). Therefore the thicker coin cell-E is subjected to four times lower C-rate than coin cell-A, and hence the $SP_{PM}$ of thicker coin cell-E would come out to be better than the thinner coin cell-A. This is because the specific performance would decrease with increasing C-rate. Hence, the cell with thicker electrodes are generally favored and must be tested in order to know the optimal cell performance.

## 4 | Implications

Table 1 depicts the normalization method which can efficiently compare the material performance and a cell performance. The existing reporting practice where the performance and current are reported by normalizing them to the mass of active cathode is a good way to compare the performance of material. Such practices were likely adopted by research community under the assumption that all the research groups maintained same electrodes thickness and similar active material loading across cells. However, in reality, the exploratory analysis is carried on different cell formats such as coin cells, swagelok cells, pouch cells, etc. In the early stages of battery research, the primary objective was to identify and develop superior active material chemistries. However, the performance of an electrochemical cell is not determined solely by the properties of the active material. In fact, there are several other factors that can influence the cell performance, many of which are overlooked when performance metrics are normalized only by the mass of the active material. Consequently, such normalization can lead to misleading interpretations of cell performance, as demonstrated in the illustration presented in Section 3.1.2. Both $SC_{PM}$ and $SC_{TM}$ increase with increasing electrode thickness up to coin cell-E before exhibiting a declining trend, whereas $SC_{AM}$ consistently decreases with increasing electrode thickness. This discrepancy highlights the limitations of active-material-based normalization. Therefore, this work focuses on developing a comprehensive framework that enables a more consistent and meaningful comparison of performance of a cell rather than just the performance of material across different studies.

**TABLE 1** | Material performance and cell performance

| S. No. | Factors | Material performance | Cell performance |
|---|---|---|---|
| 1. | Performance normalization factor ∈ (Capacity, Energy, etc.) | Mass of active material | Mass of electrochemical participants |
| 2. | Current | Density per unit gram of active material ($mAg^{-1}$) | Density per unit electrode area ($mAcm^{-2}$) |
| 3. | Cyclic charge - discharge curves data | For a given voltage window, the charge/ discharge curves must be reported over a range of areal currents ($10^{-5}mAcm^{-2}$, $10^{-3}mAcm^{-2}$, $10^{-1}mAcm^{-2}$ ...) at regular intervals of capacity retention percentage (100%, 95%, 90%, 85%, ...) | |

In case of reporting testing parameters, the current densities are reported at $mAg^{-1}$ and C-rate. The $mAg^{-1}$ captures the actual flux of electrons per mass of active material, i.e., absolute flow intensity. Comparing different material chemistries at the same $mAg^{-1}$ can unintentionally favor high-capacity materials because they are effectively operating at a lower C-rate. On the other hand, the C-rate compares the relative speed of charge/discharge of a cell in a specific time. It is a good standard to compare performance within a material, however, it is not a good standard to compare the performance across different materials. It is because, the high-capacity materials are subjected to higher current compared to small-capacity materials for a same C-rate. Comparing at the same C-rate equalizes the relative charging/ discharging demand but not the absolute current. We propose that current normalized by the area of electrode $mAcm^{-2}$ can be a good standard to compare the performance of a cell. For example, electronic current distribution, ionic current through separator, local heat generation, current collector loading, manufacturing throughput, all depend much more directly on area than on active mass. The electrochemical testing on a cell can be done at different current densities and voltages windows in order to understand the efficacy as energy storage devices or capacitive devices. It is suggestive that for a given voltage range, the electrochemical testing must be done at different areal currents such as $10^{-5}$ $mAcm{-}2$ to 1 $Acm{-}2$. The charge - discharge curves reported at regular intervals of capacity retention percentage for various areal currents can give the overall picture of a cyclic performance of a cell.

## 5 | Conclusions

With the rapid surge of interest in electrochemical energy storage technologies, several reports are published everyday on newer materials, developing synthesis and advancing cell assembly procedures with claims of outperforming the established ones. Since, the performance can vary by making slight changes in materials, synthesis, or cell assembly, it is imperative to systematically map the cause of improvement to performance and compare those efforts under one framework. However, evaluating all of those efforts gets difficult because they are not reported at a fixed cell formats. The specific performance normalized by total mass of cell ($SP_{TM}$) is desired, however, the research community reports specific performance normalized by mass of active material ($SP_{AM}$). Also, there are no reporting standards for gravimetric charge discharge curves, which delivers key performance metrics like capacity, energy density, power density, coulombic efficiency, and energy efficiency. As a result, the researchers selectively report performance which they want to highlight, and lots of data remains unreported. This study addresses the aforementioned challenges and introduces a framework designed to facilitate fair, meaningful, and efficient comparisons.

An illustration is presented in which the specific capacity, $SC_{AM}$ and $SC_{TM}$ of small-area coin cell and large-area C18650 cell are varied as a function of electrode thickness. It was observed that $SC_{AM}$ of small-area and large-area cell showed a decreasing trend with change in thickness of electrode, however their magnitudes were same across cell formats. In contrast, the $SC_{TM}$ showed an increasing trend with change in thickness of electrode, however their magnitudes differed. Therefore, the $SC_{AM}$ remain same across the cell formats but does not compare the cell performance in a way, as done by $SC_{TM}$. The $SC_{TM}$ compares the performance of a cell but it does not remain same across cell-formats making comparison difficult. To address this, we proposed a new way of reporting specific performance normalized by the mass of electrochemical participants ($SP_{PM}$). The mass of electrochemical participants includes the sum of mass of cathode, anode, their current collectors, electrolyte and separators. The $SP_{PM}$ and mass of electrochemical participants per area of the electrode can estimate the $SP_{TM}$ at any cell format - whether it be small-area laboratory cells or large-area commercial cells. Therefore, the details of component-wise masses and physical dimension of cell such as area of active electrode becomes imperative in the process of estimation. The other competencies of $SP_{PM}$ include - i) it compares the cell performance like $SP_{TM}$, ii) it can also make effective comparisons across different cell formats iii) it can be used to optimize (maximize) the cell performance as a function of thickness of electrode, loading mass of active material and also cell performance per unit price.

In this study, we have developed a comprehensive and structured framework that catalogues all the relevant information for materials used in electrochemical energy storage devices. The framework can map materials, synthesis and cell assembly to cyclic performance of a cell and can compare of all them efficiently using new specific performance, $SP_{PM}$. The framework can also suggest better combination of cell materials, promising synthesis routes to prepare them and advance techniques to assemble them into a cell within your specified conditions - whether it be in desired range of voltage window, current density, temperature, specific class of materials, or minimum cycle retention. The framework helps in summarizing the massive data of charge-discharge curves ranging upto 1000s, 10000s cycles, etc., without significant loss of information, thereby making it easier to store, interpolate and analyze. The stored charge-discharge curves in the framework along different voltage ranges help in figuring out the effective voltage window under which the cell can deliver better specific performance. The charge-discharge curves at various current density would help in distinguishing whether the material is suited for energy storage applications or capacitive applications. AI/ ML models can use structured and organized information collected in the framework to predict newer combination of materials, synthesis and assembly to its corresponding cyclic performance of a cell more accurately. Extensive information captured by the framework could be also used to develop digital twins of a cell.

The framework powered with your data would speak on your behalf regarding the performance of your cell material and where it stands in today's world of emerging class of energy storage materials. Hence, the commercial, measured and simulated data of electrochemical energy storage devices must be collected in this proposed framework. The framework can be adapted by journals to facilitate the reporting of data in a more comprehensive way.

## Author Contributions

This is an author contribution text. This is an author contribution text. This is an author contribution text. This is an author contribution text. This is an author contribution text.

## Acknowledgments

We acknowledge the Center for Atomistic Modeling and Materials Design, IIT Madras.

## Conflicts of Interest

The authors declare no conflicts of interest.

## Data Availability Statement

This is a data availability statement.

## References

1. Rares-George Scurtu, Alessandro Innocenti, Vanessa Scheck, Mario Maunz, Thomas Waldmann, Markus Hölzle, Alice Hoffmann, Peter Axmann,, and Margret Wohlfahrt-Mehrens. "From small batteries to big claims." *Nature Nanotechnology* 20, no. 7 (2025): 970–976.
2. Christian Heubner, Karsten Voigt, Peter Marcinkowski, Sebastian Reuber, Kristian Nikolowski, Michael Schneider, Mareike Partsch,, and Alexander Michaelis. "From Active Materials to Battery Cells: A Straightforward Tool to Determine Performance Metrics and Support Developments at an Application-Relevant Level." *Advanced Energy Materials* 11, no. 46 (2021): 2102647.
3. Jie Li, Catia Arbizzani, Signe Kjelstrup, Jie Xiao, Yong yao Xia, Yan Yu, Yong Yang, Ilias Belharouak, Thomas Zawodzinski, Seung-Taek Myung, Rinaldo Raccichini,, and Stefano Passerini. "Good practice guide for papers on batteries for the Journal of Power Sources." *Journal of Power Sources*452 (2020): 227824.
4. Catia Arbizzani, Yan Yu, Jie Li, Jie Xiao, Yong yao Xia, Yong Yang, Clara Santato, Rinaldo Raccichini,, and Stefano Passerini. "Good practice guide for papers on supercapacitors and related hybrid capacitors for the Journal of Power Sources." *Journal of Power Sources*450 (2020): 227636.
5. Wiley-VCH 2022. *Battery Data Reporting Checklist, https://advanced.onlinelibrary.wiley.com/pb-assets/assets/vch/msp/author_guidelines/data_reporting_checklist/Battery_Data_Reporting_Checklist-1643963769483.pdf*. Wiley-VCH.
6. Stephan, and Alexandra K.. "Standardized Battery Reporting Guidelines." *Joule* 5, no. 1 (2021): 1–2.
7. Sun, and Yang-Kook. "An Experimental Checklist for Reporting Battery Performances." *ACS Energy Letters* 6, no. 6 (2021): 2187–2189.
8. Aashutosh Mistry, Ankit Verma, Shashank Sripad, Rebecca Ciez, Valentin Sulzer, Ferran Brosa Planella, Robert Timms, Yumin Zhang, Rachel Kurchin, Philipp Dechent, Weihan Li, Samuel Greenbank, Zeeshan Ahmad, Dilip Krishnamurthy, Alexis M. Fenton Jr., Kevin Tenny, Prehit Patel, Daniel Juarez Robles, Paul Gasper, Andrew Colclasure, Artem Baskin, Corinne D. Scown, Venkat R. Subramanian, Edwin Khoo, Srikanth Allu, David Howey, Steven DeCaluwe, Scott A. Roberts,, and Venkatasubramanian Viswanathan. "A Minimal Information Set To Enable Verifiable Theoretical Battery Research." *ACS Energy Letters* 6, no. 11 (2021): 3831–3835.
9. Anubhav Jain, Shyue Ping Ong, Geoffroy Hautier, Wei Chen, William Davidson Richards, Stephen Dacek, Shreyas Cholia, Dan Gunter, David Skinner, Gerbrand Ceder,, and Kristin A. Persson. "Commentary: The Materials Project: A materials genome approach to accelerating materials innovation." *APL Materials* 1, no. 1 (2013): 011002.
10. Arunima K. Singh, Lan Zhou, Aniketa Shinde, Santosh K. Suram, Joseph H. Montoya, Donald Winston, John M. Gregoire,, and Kristin A. Persson. "Electrochemical Stability of Metastable Materials." *Chemistry of Materials* 29, no. 23 (2017): 10159–10167.
11. Jingbo Liu ed, Springer-Verlag GmbH,, and Heidelberg. "Battery-materials." *SpringerMaterials* 1, no. 1 (2024): 1.
12. Valerio De Angelis, Yuliya Preger,, and Babu R. Chalamala. "Battery Lifecycle Framework: A Flexible Repository and Visualization Tool for Battery Data from Materials Development to Field Implementation." *ECSarXiv* 1, no. 3 (2024): 1–13.
13. Tikam C. Soni, M.K. Manoj, M.L. Verma,, and Manwendra. K. Tripathi. "Supercapacitor Materials Database Generated using Web Scrapping and Natural Language Processing." *Journal of Molecular Graphics and Modelling*136 (2025): 108980.
14. Shu Huang, and Jacqueline M. Cole. "A database of battery materials auto-generated using ChemDataExtractor." *Scientific Data* 7, no. 1 (2020): 260.
15. Karishma Rawat, Anshi Pandey, Peeyush Phogat, Shreya, Ranjana jha,, and Sukhvir Singh. "Dual-ion batteries: A comprehensive review of materials, mechanisms, and market potential." *Journal of Alloys and Compounds*1031 (2025): 180827.
16. Xiangxiang Chen, Xingchen Pan, Yixin Zhao,, and Zhiyong Xie. "A li+/Na+ hybrid rechargeable full battery comprising rice husk porous hard carbon anode and sodium vanadium phosphate." *Electrochemistry Communications*172 (2025): 107877.
17. D. Lakshmi, Nithyadharseni Palaniyandy, Sivasambu Bohm,, and Bhekie B. Mamba. "Understanding the mechanisms

of mixed-ion cathode materials for aqueous and non-aqueous lithium/sodium-ion batteries  A review." *Current Opinion in Electrochemistry*38 (2023): 101217.
18. Zhiwei Chen, Xinran Li, Chaoyi Qiu, Zening Wu, Junqiang Deng, Zhichao Wang, Lingchang Wu, Haoxiang Yu, Lei Yan, Liyuan Zhang, Ting-Feng Yi,, and Jie Shu. "Expanding electrolyte window enables high voltage aqueous batteries: Challenges and strategies." *Energy Storage Materials*84 (2026): 104800.
19. Shiyi Tan, Jiaxin Meng, Yiming Wang, Guozhen Zhang, Weizhai Bao, He Liu, Cong Guo, Mingxia Song, Jingfa Li,, and Feng Yu. "Electrolyte engineering strategies towards wide voltage windows in electrical double-layer capacitors." *Journal of Power Sources*677 (2026): 239977.
20. Hong Tan, and Xiuyi Lin. "Electrolyte Design Strategies for Non-Aqueous High-Voltage Potassium-Based Batteries." *Molecules* 28, no. 2 (2023): 1–16.
21. Kelsey B. Hatzell, and Linda F. Nazar. "Recent Advances in Solid-State Batteries." *Journal of the American Chemical Society* 147, no. 18 (2025): 14881–14883.
22. Emiliano N. Primo, Matthieu Touzin,, and Alejandro A. Franco. "Calendering of Li(Ni0.33Mn0.33Co0.33)O2-Based Cathodes: Analyzing the Link Between Process Parameters and Electrode Properties by Advanced Statistics." *Batteries & Supercaps* 4, no. 5 (2021): 834–844.
23. Jiří Báňa, Pavel Čudek, Martin Šedina, Antonín Šimek,, and Tomáš Kazda. "Effect of pressing pressure on the capacity of recycled graphite anode." *Monatshefte für Chemie - Chemical Monthly* 155, no. 3 (2024): 253–259.
24. R. Borah, F.R. Hughson, J. Johnston,, and T. Nann. "On battery materials and methods." *Materials Today Advances*6 (2020): 100046.
25. Alfredo Romero-Contreras, Nora Aleyda García Gómez,, and Eduardo Sánchez-Cervantes. "Comparative study of 2D perovskite cathodes: impact of organic cation variation on electrochemical performance in lithium-ion batteries." *Journal of Electrochemical Science and Engineering* 16, no. 3234 (2026): 1–16.
26. P. Laxman Mani Kanta, N. Lakshmi Priya, Prajeet Oza, M. Venkatesh, Satyesh Kumar Yadav, Bijoy Das, G. Sundararajan,, and R. Gopalan. "Unusual Case of Higher Cyclic Stability at a Wider Voltage Window in Sodium Vanadium Phosphate." *ACS Applied Energy Materials* 4, no. 11 (2021): 12581–12592.
27. P. Laxman Mani Kanta, M. Venkatesh, Satyesh Kumar Yadav, Bijoy Das,, and R. Gopalan. "High energy-power characteristics of microstructurally engineered sodium vanadium phosphate in full cell level." *Applied Energy*334 (2023): 120665.
28. Andrey W. Golubkov, Sebastian Scheikl, René Planteu, Gernot Voitic, Helmar Wiltsche, Christoph Stangl, Gisela Fauler, Alexander Thaler,, and Viktor Hacker. "Thermal runaway of commercial 18650 Li-ion batteries with LFP and NCA cathodes - Impact of state of charge and overcharge." *RSC Advances* 5, no. 70 (2015): 57171  57186. Cited by: 512; All Open Access, Hybrid Gold Open Access.
29. Alok Kumar Mishra, and Mukul Shukla. "Effect of Electrode Thickness and Operating Temperature on Electrochemical Performance of Li-Ion Batteries." *Energy Storage* 7, no. 4 (2025): e70172. e70172 EST2-24-0684.R2.
30. Honghe Zheng, Jing Li, Xiangyun Song, Gao Liu,, and Vincent S. Battaglia. "A comprehensive understanding of electrode thickness effects on the electrochemical performances of Li-ion battery cathodes." *Electrochimica Acta*71 (2012): 258–265.
31. Junsheng Zheng, Guangguang Xing, Liming Jin, Yanyan Lu, Nan Qin, Shansong Gao,, and Jim P. Zheng. "Strategies and Challenge of Thick Electrodes for Energy Storage: A Review." *Batteries* 9, no. 3 (2023): 1–21.
32. Han Gao, Qiang Wu, Yixin Hu, Jim P. Zheng, Khalil Amine,, and Zonghai Chen. "Revealing the Rate-Limiting Li-Ion Diffusion Pathway in Ultrathick Electrodes for Li-Ion Batteries." *The Journal of Physical Chemistry Letters* 9, no. 17 (2018): 5100–5104. PMID: 30130117.
33. Yujin Kim, Moonjin Kim, Taeyong Lee, Eunchae Kim, Minju An, Joohyuk Park, Jaephil Cho,, and Yeonguk Son. "Investigation of mass loading of cathode materials for high energy lithium-ion batteries." *Electrochemistry Communications*147 (2023): 107437.
34. J. Mitali, S. Dhinakaran,, and A.A. Mohamad. "Energy storage systems: a review." *Energy Storage and Saving* 1, no. 3 (2022): 166–216.
35. Muhammad Zahir Iqbal, Mian Muhammad Faisal,, and Syeda Ramsha Ali. "Integration of supercapacitors and batteries towards high-performance hybrid energy storage devices." *International Journal of Energy Research* 45, no. 2 (2021): 1449–1479.
36. Mukrimin Sevket Guney, and Yalcin Tepe. "Classification and assessment of energy storage systems." *Renewable and Sustainable Energy Reviews*75 (2017): 1187–1197.
37. Mohanraj Kandhasamy, Barani Kumar Duvaragan, Santhosh Kamaraj,, and Ganesan Shanmugam. "An Overview on Classification of Energy Storage Systems." *Advances in Sustainable Energy Technologies*1484 (2024): 1–25.
38. F.J. Rodríguez-Varela, Ivonne Liliana Alonso Lemus, Oumarou Savadogo,, and Karthikeyan Palaniswamy. "Overview: Current trends in green electrochemical energy conversion and storage." *Journal of Materials Research*36 (2021): 40714083.
39. Sandeep A Arote 2021. "Electrochemical energy storage mechanisms and performance assessments: an overview." In *Electrochemical Energy Storage Devices and Supercapacitors*, 2053-2563, 1–1 to 1–34. IOP Publishing.
40. Dongjian Li, Qiqi Lv, Chunmei Zhang, Wei Zhou, Hongtao Guo, Shaohua Jiang,, and Zhuan Li. "The Effect of Electrode Thickness on the High-Current Discharge and Long-Term Cycle Performance of a Lithium-Ion Battery." *Batteries* 8, no. 8 (2022): 1–13.
41. Yige Xiong, Die Zhang, Xiaorong Ruan, Shanbao Jiang, Xueqin Zou, Wei Yuan, Xiuxue Liu, Yapeng Zhang, Zeqi Nie, Donghai Wei, Yubin Zeng, Peng Cao,, and Guanhua Zhang. "Artificial intelligence in rechargeable battery: Advancements and prospects." *Energy Storage Materials*73 (2024): 103860.
42. Muntather Almusawi, Aasheesh Shukla, Hemalatha S, P. Kavitha, G.M. Gambhire, Pankaj R. Pardeshi,, and B. Pragathi. "Comparative Analysis of Supercapacitors vs. Batteries." *E3S Web Conf.*591 (2024): 01010.
43. Tholkappiyan Ramachandran, Siva Sankar Sana, Kulurumotlakatla Dasha Kumar, Yedluri Anil Kumar, H.H. Hegazy,, and Seong Cheol Kim. "Asymmetric supercapacitors: Unlocking the energy storage revolution." *Journal of Energy Storage*73 (2023): 109096.
44. Kihyun Son, Soo Min Hwang, Seok-Gyun Woo, Jin Kyo Koo, Meenseon Paik, Eui Hwan Song,, and Young-Jun Kim. "Comparative study of thermal runaway and cell failure of lab-scale Li-ion batteries using accelerating rate calorimetry." *Journal of Industrial and Engineering Chemistry*83 (2020): 247–251.
45. Dhrubajyoti Bhattacharjya, Daniel Carriazo, Jon Ajuria,, and Aitor Villaverde. "Study of electrode processing and cell assembly for the optimized performance of supercapacitor in pouch cell configuration." *Journal of Power Sources*439 (2019): 227106.
46. Anna Smith, Pirmin Stüble, Lea Leuthner, Andreas Hofmann, Fabian Jeschull,, and Liuda Mereacre. "Potential and Limitations of Research Battery Cell Types for Electrochemical Data Acquisition." *Batteries & Supercaps* 6, no. 6 (2023): e202300080.
47. Zhan Lin, Tiefeng Liu, Xinping Ai,, and Chengdu Liang. "Aligning academia and industry for unified battery performance metrics." *Nature Communications* 9, no. 1 (2018): 5262.
48. Kevin W. Knehr, Joseph J. Kubal, Abhas Deva, Mohammed B. Effat,, and Shabbir Ahmed. "From material properties to device metrics: a data-driven guide to battery designElectronic supplementary information (ESI) available. See DOI: https://doi.org/10.1039/d3ya00137g." *Energy Advances* 2, no. 9 (2023): 1326–1350.

## Supporting Information

Supplementary data associated with this article can be found in the drive link https://drive.google.com/drive/folders/1z7kczMcWz7BBls9kfxkhezI4E4mfNZTG?usp=sharing. The details of component-wise masses of coin cells used for illustration is found in this PDF file. An Excel sheet containing a database format. Another excel sheet, where Tattvasar calculator is incorporated.